\documentclass{desyproc}

\begin{document}
\title{STAX. An Axion-like Particle Search with Microwave Photons}

\author{{\slshape  J. Ferretti$^{1}$}\\[1ex]
$^1$Dipartimento di Fisica and INFN, `Sapienza' Universit\`a di Roma, P.le Aldo Moro 5, 00185 Roma, Italy}

\contribID{familyname\_firstname}

\confID{13889}  
\desyproc{DESY-PROC-2016-XX}
\acronym{Patras 2016} 
\doi  

\maketitle

\begin{abstract}
We discuss an improved detection scheme for a light-shining-through-wall (LSW) experiment for axion-like particle searches. We propose to use: gyrotrons or klystrons, which can provide extremely intense photon fluxes at frequencies around 30 GHz; transition-edge-sensors (TES) single photon detectors in this frequency domain, with efficiency $\approx1$; high quality factor Fabry-Perot cavities in the microwave domain, both on the photon-axion conversion and photon regeneration sides.
We compute that present laboratory exclusion limits on axion-like particles might be improved by at least four orders of magnitude for axion masses $\lesssim 0.02$ meV.
\end{abstract}

\section{Introduction}
Axions \cite{gen} are between the most serious dark matter candidates. They are light neutral scalar or pseudoscalar bosons, with mass $m_a \approx \mu$eV$-m$eV, coupled to the electromagnetic field via
\begin{equation}
	\mathcal{L}_I=\frac{1}{4}G\,a\,F^{\mu\nu}\tilde{F}_{\mu\nu}
	\label{elle}
\end{equation}
In QCD axion models (DFSZ \cite{dfsz} and KSVZ \cite{ksvz}), the axion-photon coupling constant $G$ is directly related to $m_a$; thus, $G$ is the only free parameter of the theory. 
In axion-like particle (ALP) searches, the parameter space is extended: $G$ and $m_a$ are the free parameters \cite{pdg}.

Axions and ALPs experimental searches can be divided into two main categories: 1) Axions from astrophysical and cosmological sources; 2) Laboratory searches.
In the former case, exclusion limits on the axion-photon coupling constant are provided by estimates of stellar-energy losses \cite{pdg,en-losses}, helioscopes \cite{sikivie:1983,Arik:2015cjv,iaxo} and haloscopes \cite{sikivie:1983,admx}. In the latter case, limits on $G$ are given by photon polarization \cite{pol-exp} and Light-Shining-Through-Wall (LSW) \cite{sikivie:1983,ALPS,Betz:2013dza,alps2,oscar} experiments. 

In this contribution, we will focus on LSW experiments. After a brief description of the standard LSW experimental apparatus, we will discuss how to improve present ALPs laboratory limits on $G$ by at least four orders of magnitude \cite{stax}. We are willing to do this by using extremely intense photon fluxes from gyrotron sources at frequencies around 30 GHz, TES single photon detectors with efficiency $\approx1$, and high quality factor Fabry-Perot cavities in the microwave domain ($Q \approx 10^4-10^5$), both on the photon-axion conversion and photon regeneration sides.



\begin{figure}[htbp]
\centerline{\includegraphics[width=0.6\textwidth]{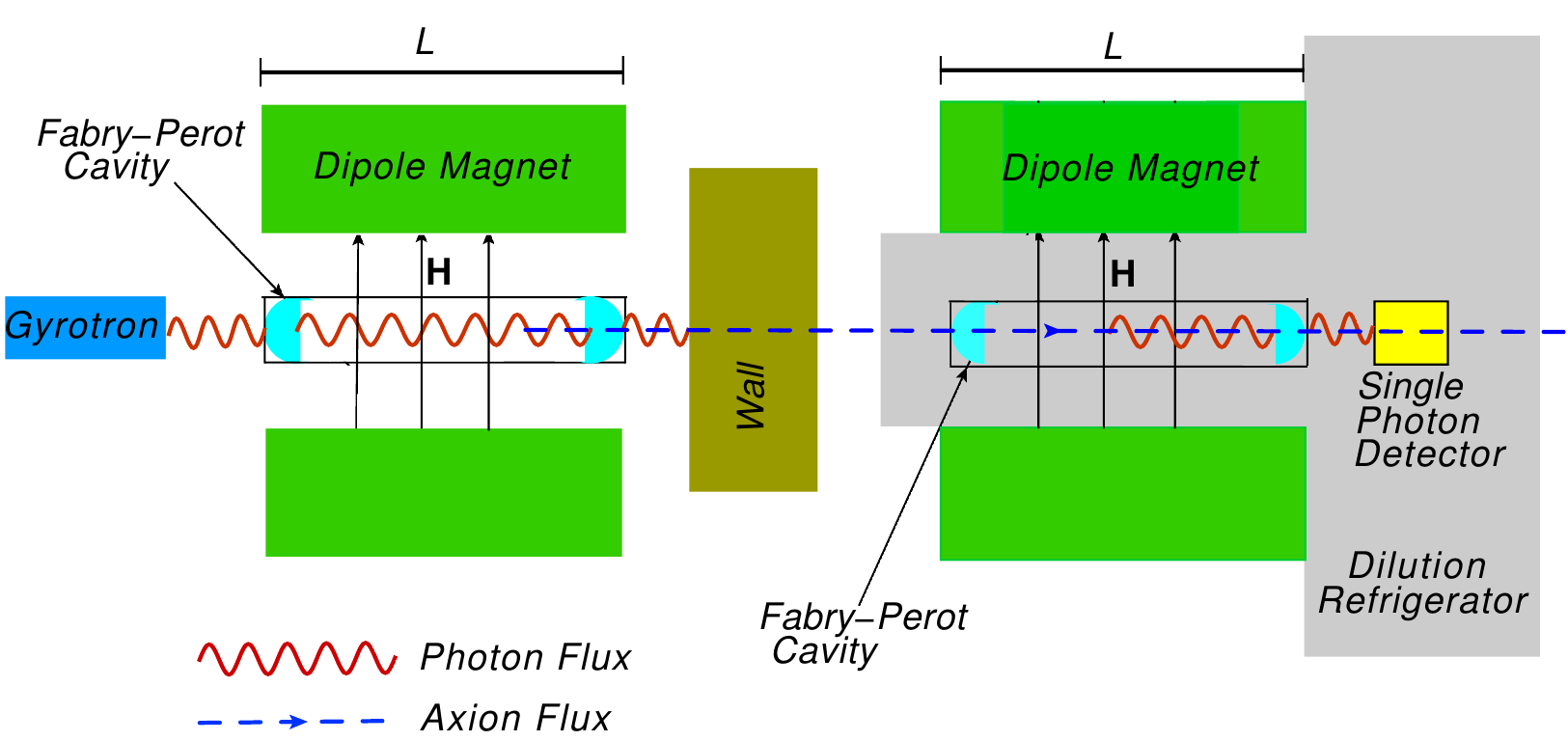}}
\caption{Experimental configuration of the STAX LSW experiment. Fig. from Ref. \cite{stax}; Elsevier B.V. copyright.}
\label{Fig:uno}
\end{figure}

\section{LSW experiments}
In a LSW experiment \cite{sikivie:1983,ALPS,Betz:2013dza,alps2,oscar}, a coherent photon beam traverses an intense magnetic field, $\bf H$.
Here, some photons can convert into axions via the Primakoff effect. 
Photons exchange 3-momentum $\bf q$ with $\bf H$, the energy is conserved. 
If the $\hat x$ axis is chosen in the direction of the propagating photon beam, then the external magnetic is assumed to be uniform in the volume $L_x L_y L_z = L_x S$.

The photons which do not convert into axions are stopped by an optical barrier, ``the wall". while axions can cross the wall, due to their negligible cross-section with ordinary matter. On the other side of the wall there is a second magnetic field, which can convert axions back to photons. Reconverted photons may be detected via a single-photon detector. 

In the $\epsilon_\gamma \gg m_a$ limit, the photon to axion (axion to photon) conversion probability is given by \cite{sikivie:1983}
\begin{equation}
	P_{\gamma \rightarrow a} = P_{a \rightarrow \gamma} = G^2 H^2 \mbox{ } \frac{\sin^2(q_xL_x/2)}{q_x^2} \mbox{ } \frac{\epsilon_\gamma}{\sqrt{\epsilon_\gamma^2 - m_a^2}}
\end{equation}
where $\epsilon_\gamma$ is the photon (axion) energy and $m_a$ the axion mass.
In the limit $\epsilon_\gamma \approx m_a$, which is relevant for the STAX experiment, the previous expression for the conversion probability has to be regulated \cite{stax}
\begin{equation}
	P_{\gamma \rightarrow a} = P_{a \rightarrow \gamma} = G^2 H^2 \mbox{ } \frac{\sin^2(q_xL_x/2)}{q_x^2} \mbox{ } 
	\frac{\epsilon_\gamma}{\frac{1}{L_x} + \sqrt{\epsilon_\gamma^2 - m_a^2}}
\end{equation}
The photon-axion-photon rate reads
\begin{equation}
	\frac{{\rm d}N_\gamma}{{\rm d} t} = \Phi_\gamma \eta P_{\gamma \rightarrow a}^2
\end{equation}
where $\Phi_\gamma$ [s$^{-1}$] is the initial photon flux and $\eta$ the single-photon-detector efficiency. The rate can be increased by introducing a Fabry-Perot cavity in the magnetic field area before the wall by a factor of $Q$, which is the quality factor of the cavity. Moreover, as discussed in Ref. \cite{Sikivie:2007}, the rate can be further increased with the addition of a second Fabry-Perot cavity in the magnetic field region beyond the wall. See Fig. \ref{Fig:uno}.

\section{STAX experimental configuration and calculated exclusion limits}
The best laboratory limits for the axion-photon coupling constant have been provided by the ALPS Collaboration \cite{ALPS}. The second stage of ALPS, ALPS-II \cite{alps2}, will improve the previous limits mainly by increasing the magnetic field length as well as introducing a second cavity in the magnetic field region behind the wall. ALPS-II configuration is very similar to that of Fig. \ref{Fig:uno}, but in this case the photon flux is provided by an optical laser.
\begin{figure}[htbp]
\centerline{\includegraphics[width=0.5\textwidth]{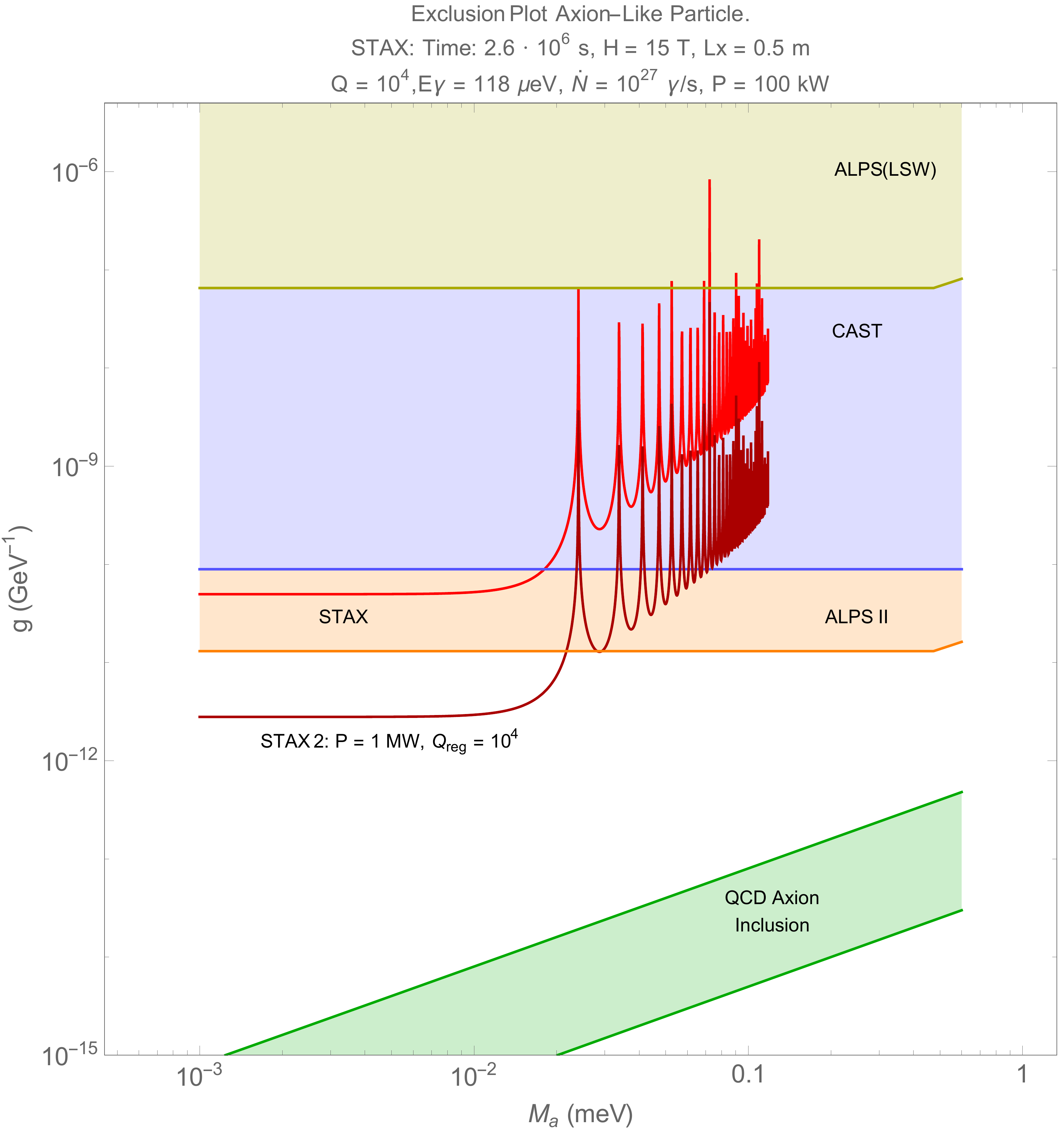}}
\caption{$90\%$ CL exclusion limits that STAX and STAX 2 may achieve in case of a null result for axions with $m_a \lesssim 0.02$ meV. An exposure time of one month and zero dark counts are considered. ``STAX" and ``STAX 2" configurations correspond to a 100 kW and 1 MW gyrotron sources, respectively. Picture from Ref. \cite{stax}; Elsevier B.V. copyright.}
\label{Fig:MV}
\end{figure}

Our goal is to develop a new generation LSW experiment and improve the limits on $G$ by using sub-THz photon sources.
Sub-THz sources, like gyrotrons and klystrons, can provide very high powers (up to 1 MW) at small photon frequencies, resulting in photon fluxes up to $10^{10}$ more intense than those from optical lasers, used in previous LSW experiments. 
We will also use high Q-factor Fabry-Perot cavities for microwave photons and single-photon detectors for light at these frequencies, with almost zero dark count, based on the (Transition-Edge-Sensor) TES technology.
The TES detector will be  coupled to an antenna and operated at temperatures $\approx 10$ mK.

In this way, we computed that present laboratory exclusion limits on axion-like particles might be improved by at least four orders of magnitude for axion masses $\lesssim 0.02$ meV \cite{stax}. The limits that STAX experiment may achieve are compared to previous experimental results in Fig. \ref{Fig:MV}.

\begin{footnotesize}

\end{footnotesize}


\end{document}